# Formalization and Verification of Hierarchical Use of Interaction Overview Diagrams Using Timing Diagrams

Aymen Louati[1,3]

[1]LR-SITI, ENIT, Université Tunis
El Manar, BP 37, Tunis, Tunisie
aymen.louati@enit.rnu.tn

Chadlia Jerad[2]

[2]OASIS, ENIT, Université Tunis
El Manar, BP 37, Tunis, Tunisie
chadlia.jerad@gmail.com

Kamel Barkaoui[3]

[3]CEDRIC–CNAM, 292, Rue
Saint-Martin, Paris, France
kamel.barkaoui@cnam.fr

*Abstract*—Thanks to its graphical notation and simplicity, Unified Modeling Language (UML) is a de facto standard and a widespread language used in both industry and academia, despite the fact that its semantics is still informal. The Interaction Overview Diagram (IOD) is introduced in UML2; it allows the specification of the behavior in the hierarchical way. In this paper, we make a contribution towards a formal dynamic semantics of UML2. We start by formalizing the Hierarchical use of IOD. Afterward, we complete the mapping of IOD, Sequence Diagrams and Timing Diagrams into Hierarchical Colored Petri Nets (HCPNs) using the Timed colored Petri Nets (timed CP-net). Our approach helps designers to get benefits from abstraction as well as refinement at more than two levels of hierarchy which reduces verification complexity.

## I. INTRODUCTION

Nowadays, UML is the most adopted semi-formal language for system modeling [16]. Despite his widespread use, users do agree on the interpretation of only few well-known concepts, while precise meaning of many parts of the notation is still missing. The migration of UML1 to UML2 brought more precision. Nevertheless, latter remains informal and lacks tools for automatic analysis and validation. Since a major challenge in software development process is to advance error detection to early phases of the software life cycle. Many works [7],[8] dealt with their formalization, they tried to combine the simple and ease of use of UML with the reasoning and analysis capabilities of formal methods.

UML2 introduced a new diagram, which is the IOD[1]. The main purpose of the IOD is to show the interaction of the components within the system at high level of abstraction. It is derived from UML2 activities that can only have interaction elements or interaction uses instead of invocation operations. Several reasons explain the need to use IOD in a hierarchical way. Firstly, it's not practical to draw the behavior of very large system with a single diagram. Secondly, it can be seen as black boxes allowing the modeler to work at different abstraction levels and by using different refinement techniques. The goal of our work is to provide formalization of hierarchical use of IOD semantics into terms of HCPNs, where one of the branches may be represented by UML2 Timing Diagram (TD for short), UML2 Sequence Diagram (SD for short), or IOD.

In order to formalize this hierarchical use, HCPNs appear to be suitable for this purpose, due to their structure. Our work is an extension of Tebibel's studies [1],[2],[3] with more than two levels of hierarchy. So, we propose to use the timed CP-net in our approach, for formalizing TD. The remainder of our paper starts with an overview on related work focuses on formal verification of UML2 specifications. In Section III, we present the formal definition of hierarchical use of IOD. The hierarchical mapping of all IOD constructs into HCPNs and timed CP-net is presented in section IV and illustrated through a case study in section V. Finally, we drawn in section VI, a conclusion and announces our future work.

## II. RELATED WORK

In literature, several works dealt with the validation of structural [14], as well as behavioral [3], [4], [5], [7], [11], [16] ,[17] UML diagrams, or even both [6],[15].
The first attempt to formalize UML2 activities was introduced by Störrle in [7], [8], [9], [10], where he used the colored Petri nets (CPN).

In [5], the authors treat with consistency checking of UML behavioral diagrams by Petri nets (PN). Although, IOD plays key role for the description of components interactions, only few work deal with their formalization. In this work, we are particularly interested on formalization and verification of the hierarchical use of IOD. Indeed, this diagram despite its importance, we find Tebibel's studies [1],[2],[3], Baresi and all studies [6] and Andrade and all studies [16],[17].
In [6], the authors propose a formal verification of timed systems by using the MADES modeling notation, borrowing many concepts from SysML[2] and MARTE[3] for describing temporal notation. They allow checking temporal properties. Also in [16], [17], MARTE and SysML are used for mapping IOD and activities into a Time PN with energy constraints. The authors tried to present a formal verification of Embedded and Real-Time Systems. The first works proposing an approach for mapping IOD into HCPNs were [1], [2], [3]. We also find in [12], the translation of OCL invariants into CTL formulas in

---

order to check the properties within the HCPN. Despite the relevance of the results, both works use IOD for modeling interactions between components, but not in a hierarchical way. That is, the interaction nodes of the main IOD refer only to SD. However these nodes may refer to others interaction diagrams such as communication diagrams, TD or even others IOD.

This characteristic improves the expressiveness of the model. For filling this lack, we try to propose a new approach by extending Tebibel's studies for giving a formal description in hierarchical way at more than two levels. So, we give the rules and algorithms of translation basis IOD elements with an SD and TD models into terms of HCPNs using the timed CP-net for mapping TD. So, the models are consistent before and after conversion.

## III. FORMAL DEFINITION OF IOD, SD AND TD

### A. UML2 Interaction Overview Diagrams

We define in first the IOD. It's a special type of UML2 activities where nodes can refer interaction occurrences (or interaction use) **Fig1.a** or interaction elements (or interaction) **Fig1.b**. They mean respectively a reference to existing interaction diagrams and display a representation of existing interaction diagrams. IOD, SD, and TD are particular cases of UML2 interaction. This formalism allows a valuated control flow specification in hierarchical way and takes the same UML2 activity diagram notations (initial, final, join, fork nodes etc.). We start by recalling the work of Tebibel and all, where they show the interaction between system components using only one IOD and a set of SD such as interaction nodes. This meaning that, the interaction nodes of the IOD refer only and exactly to SD. For this purpose, they proposed the following formalizations of IOD and SD. In [3], the authors propose a formal definition of IOD by the n-tuple as follows:

$M_{IOD} = (n_0, NF, I, B, D, E, Ed)$ where:

- $n_0$ is the initial node.
- $NF = (nf1,..,nf_n)$ is a finite set of final nodes.
- $I = (in1,..,in_n)$ is a finite set of inodes.
- $B = (b_1,..,b_n)$ is a finite set of join and fork nodes.
- $D = (d_1,..,d_n)$ is a finite set of decision and merge nodes.
- $E = (e_1,..,e_n)$ is a finite set of edges connecting IOD nodes.
- $Ed = \{n_0\} \cup I \cup B \cup D \times NF \cup I \cup B \cup D \rightarrow E$ is a function which connects IOD nodes by edges.

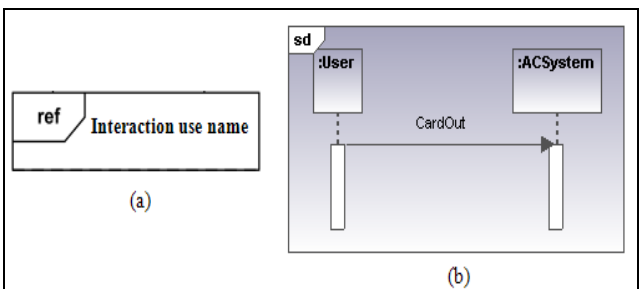

Fig1. a. interaction use   b. interaction element now brings together various information, messages, and objects for describing interactions involved on the sequencing of messages exchanged between objects and represented by life lines. In [3], we keep the formal definition of this diagram by the n-tuple as follows:

$M_{SD} = $ (Lf, Msg, Beg, End, Ptx, Find, Lost, Alt, Op, Par, Loop, In, Out, Str) where:

- $Lf = \{lf_1,..,lf_n\}$ is a finite set of lifelines.
- $Msg = \{msg_1,..,msg_n\}$ is a finite set of exchanged messages between lifelines.
- $Beg = \{beg_1,..,beg_n\}$ is a finite set of interaction points on a lifeline, starting messages.
- $End = \{end_1,..,end_n\}$ is a finite set of interaction points on a lifeline, ending asynchronous messages.
- $Ptx = \{ptx_1,..,ptx_n\}$ is a finite set of interaction points on a lifeline, ending synchronous messages.
- $Find = \{f_1,..,f_n\} \subset Msg$ is a subset of all founded messages.
- $Lost = \{l_1,..,l_n\} \subset Msg$ is a subset of all lost messages.
- $Alt = \{alt_1,..,alt_n\}$ is a finite set of alternative interaction nodes.
- $Op = \{op_1,..,op_n\}$ is a finite set of optional interaction nodes.
- $Par = \{par_1,..,par_n\}$ is a finite set of parallel interaction nodes.
- $Loop = \{loop_1,..,loop_n\}$ is a finite set of loop interaction nodes.
- $In: Msg \rightarrow Beg$ is a function witch returns the interaction point to the output of a message on a lifeline.
- $Out: Msg \rightarrow End$ is a function witch returns the interaction point at the entrance of a message on a lifeline.

Since IOD is natively hierarchical, its formalization should highlight its hierarchical nature. In order to set our approach for mapping them to HCPN, we start by formalizing the hierarchical use of IOD in a way similar to HCPN. So, we reformulate the formal description of IOD to their hierarchical nature. In the rest of the paper, we consider an UML model M composed of a set of IOD, a set of SD and set of TD all related hierarchically. Thus, $M = M_{IOD} \cup M_{SD} \cup M_{TD}$ where $M_{IOD}$ represent a finite set of IOD, $M_{SD}$ represents a finite set of SD, and $M_{TD}$ represents a finite set of TD. First, we formally define a set of hierarchical IODs by the n-tuple as follows:

$M_{IOD} = (S_{IOD}, Ni, NF, I, B, D, IODcomp, E, Ed, Ref)$ where:

- $S_{IOD} = (IOD_0,..,IOD_i)$ is a finite set of IOD occurrences.
- $Ni = (ni_1,..,ni_i)$ is a finite set of all initial nodes.
- $NF = (nf_1,..,nf_i)$ is a finite set of all final nodes.
- $I = (in_1,..,in_i)$ is a finite set of all interaction nodes.

- B = $(b_1,..,b_i)$ is a finite set of all join and fork nodes.

- D = $(d1,..,di)$ is a finite set of all decision and merge nodes.

- E = $(e1,..,ei)$ is a finite set of edges connecting IOD nodes.

- IODcomp:SIOD→partition(Ni $\cup$ NF $\cup$ I $\cup$ B $\cup$ D) is a function which associates to each IOD its initial node, final nodes, its subset of interaction nodes, join and fork nodes and decision and merge nodes in level n, (n>0).

- Ed $\subset$ Ni $\cup$ I $\cup$ B $\cup$ D x NF $\cup$ I $\cup$ B $\cup$ D is an application which connects IOD nodes by edges.

- Ref: I→$M_{IOD}$ $\cup$ $M_{TD}$ $\cup$ $M_{TD}$ is an injective function which associates to each interaction nodes, the corresponding $M_{IOD}$, $M_{TD}$ or $M_{TD}$ such as there exists one and only one IOD $IOD_i$ such that $Ref^{-1}$ ($IOD_i$) = $\emptyset$.

In the model M, each interaction node references a diagram. This can be another IOD, an SD or a TD. In other terms, each interaction node should be mapped to exactly an element using the function Ref. This function captures the hierarchical structure of IOD, by associating to each interaction node its referenced diagram. If this last is an IOD, then the function IODComp returns the set nodes composing it. Starting from this definition, we deduce that Ref is injective. Since the main IOD is not referenced by any node, the image and the co-domain of Ref are not equal. Ref is also not surjective.

In order to redefine the mapping function that transforms an UML Model M consisting of IODs, SD and TDs into a HCPN $M_{HCPN}$, we need to define diagrams and interaction nodes hierarchy level. In recursive way, we define the hierarchical level n, n $\in$ N, of an IOD by the definition 3.1:

*Definition 3.1:* an IOD $IOD_j$ is hierarchical level n, such that n>0, if and only if $Ref^{-1}(IOD_j)$ belongs to an IOD of hierarchical level n-1. The IOD $IOD_j$ such that $Ref^{-1}(IOD_j)$ =$\emptyset$ is called of Hierarchical level 0.

We derive the hierarchical level of an interaction node, this is illustrates by the definition 3.2:

*Definition3.2:* an interaction node in is of hierarchical level n (n∈N), if and only if, it belongs to an IOD of hierarchical level n.

Now, we can define the hierarchical level of SD and TD respectively illustrates by the definitions 3.3 and 3.4:

*Definition 3.3:* an SD $SD_j$ is of hierarchical level n, if and only if, $Ref^{-1}(SD_j)$ belongs to an IOD of hierarchical level n-1.
*Definition 3.4:* a TD $TD_j$ is of hierarchical level n, if and only if, $Ref^{-1}(TD_j)$ belongs to an IOD of hierarchical level n-1.

### B. Timing diagram (TD)

TD is a new formalism provides by UML. It derived from techniques known system engineering and interaction diagrams. It merges state machine and SD for showing the evolution of the state of an object in the time and messages that modify this state. The appropriate elements are as follows [18]: Life line: represents an individual participant in the interaction; State or condition: represents the state of the classifier or attribute, or some testable condition; Duration constraint: constrains the time that a lifeline can maintain a state; Time constraint: constrains the time when the state transition can occur; Event: represents the trigger of transition; Message: represents an asynchronous message, and a call and a reply. In our work, we used TD when an element of IOD branch is reason about time; we will also propose its translation basing a timed CP-net. In our approach, we propose the formal definition of TD by the n-tuple as follows:

$M_{TD}$= (LF, PT, MSG, STATE, DC, TC, Event) where:

- LF= $(lf_1,..,lf_n)$ is a finite set of lifelines.

- PT= $(pt_1,..,pt_n)$ is a finite set of interactions points between lifelines and asynchronous messages.

- MSG= $(msg_1,..,msg_n)$ is a finite set of asynchronous messages exchanged between objects.

- STATE= $(st_1,..,st_n)$ is a finite set of state objects.

- DC= $(dc_1,..,dc_n)$ is a finite set of all duration constraints when a lifeline can maintains a state.

- TC= $(tc_1,..,tc_n)$ is a finite set of all time where the state transition occur.

- Event= $(e_1,..,e_n)$ is a finite set of all trigger of all transitions.

## IV. FROM THE HIERARCHICAL USE OF IOD TO HCPNs USING THE TIMED CP-net

In our case, the choice of HCPN is fully justified. We first start by introducing their formalization. Nets similar to modular programming, the construction of CPNs can be broken into smaller pieces by utilizing substitution transitions. Conceptually, nets with substitution transitions are nets with multiple layers of detail. A simplified net gives a broad overview of the system and by substituting transitions of this top-level net with sub-nets, more details could be brought into the model. HCPN as formalized by Jensen in [13], and implemented in CPN Tools, introduce a facility for building PN out of sub-nets. Also, it makes possible to edit, simulate and analyze PN models. Consequently, the idea behind HCPN theory is to allow the construction of a large model by using a number of small PNs, which are related to each other in a well-defined way. We recall the work of [3] where HCPN is defined and we propose a formal definition of HCPNs called $M_{HCPN}$. We define this by the n-tuple as follows:

$M_{HCPN}$= (Pg, P, T, SubTr, A, C, Pre, Post, Pl, Trs, Trsub, TrPg, $M_0$) where:

- Pg= ($pg_0$, $pg_1$,..., $pg_i$) is a finite set of pages, where $pg_0$ is the prime page.

- P= ($p_0$,$p_1$,...,$p_i$)is a finite set of all places.

- T= Ts ∪ SubTr = ($t_0$,$t_1$,...,$t_i$) is a finite set of all transitions disjoint of P (P∩T=∅) and where:

    - Ts= ($ts_0$,$ts_1$,...,$ts_i$) is a set of all ordinary transitions.

    - SubTr= ($subTr_0$, $subTr_1$,...,$subTr_i$) is a set of all substitution transitions.

- A⊂ P x T ∪ T x P is a finite set of all arcs.

- C= ($c_1$,...,$c_i$) is a set of colors defining tokens.

- Pre=P x T→partition(C) is the precondition to the transition firing such that Pre ($pi$,$tj$)=($c_1$,$c_2$,$c_3$,..,$c_k$).

- Post=T x P→partition(C) is the post condition to the transition firing such that Post ($ti$,$pj$)=($c_1$,$c_2$,$c_3$,..,$c_k$).

- Pl: Pg→partition (P) is a function which yields the set of places of a page.

- Trs: pg→partition(Ts) is a function which yields the set of ordinary transitions of a page.

- Trsub: pg→partition(Subtr) is a function which yields the set of substitution transitions of a page.

- TrPg(Subtr,pg) is a function which associates a page to a substitution transition.

- And $M_0$: P→C is the initial marking function, such that $M_0$($pi$) = Σ $c_k$ , k=(1,...,i).

In **[3]**, the author opted for formalizing IOD using HCPN. The choice is obvious, since this last supports perfectly hierarchical modeling. In the proposed approach, the IOD is mapped to a HCPN prime page and the interaction nodes to HCPN pages abstracted by means of substitution transitions. When creating a page, it is equipped with input and output parameters. The sub-net derived from the SD shows the end of the branch of hierarchical IOD and it is connected to these parameters. Each of these pages begins and end by transitions respectively called In-transition and Out-transition, readers can see [3] for more details. Also, we propose a formal definition of timed CP-net called $M_{TCPN}$ for transforming TD. We define this by the n-tuple as follows:

$M_{TCPN}$= (P, T, A, Σ, C, G, E, If) where:

- P= ($p_0$,$p_1$,...,$p_i$) is a finite set of places.

- T= ($t_0$,$t_1$,...,$t_i$) is a finite set of transitions such that (P∩T=∅).

- A⊂ P x T ∪ T x P is a set of all direct arcs.

- Σ is a finite set of no-empty color sets, each color set is timed.

- C:P→ Σ is a color set function that assigns a color set to each place, a place p is timed if C(p) is timed.

- G:T→ExpGF is a guard function that assigns a guard to each transition t.

- Temp=($temp_1$,...,$temp_i$) is a finite set of all time execution where transition occurs.

- E:A→EXP is an arc expression function that assigns an arc expression to each arc a that type[E(a)]=C(p), p is timed and connected to arc a.

- And If:A→EXP is an initialization expression to each place that Type[If(p)]=C(p), and p is timed.

In order to formalize the mapping, the authors defined a function Ω that transforms a given IOD into a HCPN. This function is defined by the equation (1) as follows:

$$\Omega: \{n_0\} \cup N_f \cup B \cup D \cup I \cup E \rightarrow partition\ (Pg \cup P \cup Ts \cup SubTr \cup A \cup C) \qquad (1)$$

The function Ω is no more applicable when the model of a system is composed of a set of hierarchical IOD and a set of SD. This does not mean that we have to redefine the function from scratch, but we should bring some modifications. In the following, we will present the new function $\Omega_H$ that is highly inspired from the function Ω.

In our approach, an IOD is not directly transformed into a HCPN prime page. It is only the IOD of hierarchical level 0 that is transformed. So, all the other IODs, that are of hierarchical level n such that n>0, are transformed into HCPN pages. All the SDs and TDs of the model is transformed into pages of hierarchical level n. Consequently, the function takes into account the hierarchical level of the diagram. Formally, we define $\Omega_H$ by the equation (2) as follows:

$$\Omega_H: S_{IOD} \cup Ni \cup Nf \cup B \cup D \cup I \cup E \rightarrow partition\ (Pg \cup P \cup Ts \cup SubTr \cup A \cup C) \qquad (2)$$

Table I shows the transformation of the hierarchical IOD constructs into HCPNs. For each IOD construct, we find the equivalent HCPN construct expressed by an intuitive transformation on as well as a more formal transformation. The table (Table II) yields more details on the IOD edges mapping. Each table line shows the transformation of an edge set between input and output nodes. The terms Initial, interaction and final correspond to such nodes. Transitions derived from join and fork nodes or edges are fired with respect to pre and post conditions. The considered model is composed of IODs and SDs. The sub-nets mapping the IODs are obtained by applying the function $\Omega_H$. However, the sub-nets mapping the SDs result from the application of the function introduced in [3] illustrated by the equation (3), also, the sub-nets mapping the TDs that we propose in our work is a function illustrated by the equation (4):

$$\Phi{:}Lf \cup Msg \cup Beg \cup End \cup Ptx \cup Find \cup Lost \cup Alt \cup Op \cup Par \cup \cup Loop \rightarrow partition \ (Pg \cup P \cup Ts \cup SubTr \cup A \cup C \cup \ Pr) \quad (3)$$

$$\theta{:}LF \cup MSG \cup STATE \cup DC \cup TC \cup Event \rightarrow partition \ (Pg \cup P \cup T \cup SubTr \cup A \cup \Sigma \cup C \cup G \cup Temp \cup E \cup If) \quad (4)$$

The function $\Phi$ is kept as it is with no changes, except the description of Pl, Trs and Trsub functions update with the additional places, transitions and substitution transitions. Readers may refer to [2] for further details. We propose to add the function $\theta$ that illustrates the rule transformation of TD constructs to timed CP-net. All translation rules are presented on table III.

Table I.   Mapping of Hierarchical use of IOD

| Rule | IOD constructs | HCPN | |
|------|----------------|------|---|
| | | $\Omega_H$ (translation rules) | Intuitive translation |
| 1 | $IOD_n$ of hierarchical level 0 | If $IOD_n$ and level (IOD) = 0 then Create Page(pg$_0$); | prime page |
| 2 | $IOD_n$ of hierarchical level n, n>0 | If $IOD_n$ and level(IOD)>0 then Create Page(pg$_i$) ∈ Pg; | page |
| 3 | an interaction node in on level n (Inline Interaction) | ∀ in ∈ I in level n, Create a substitution transition subtr$_i$ ∈ Subtr in pg$_n$; Create a page pg$_i$ ∈ Pg$_{n+1}$, $\Omega_H$(Ref(in$_i$))=subtr$_i$; Trpg(subtr$_i$,pg$_i$); Trsub($\Omega_H$(IODcomp$^{-1}$(ini))) = Trsub($\Omega_H$(IODcomp$^{-1}$(ini))) ∪ subtr$_i$; | substitution transition |
| 4 | initial node ni level n | ∀ ni$_i$ ∈ Ni in level n, Create a place p$_i$ ∈ pg; Pl($\Omega_H$(IODcomp$^{-1}$(ni$_i$))) = Pl($\Omega_H$(IODcomp$^{-1}$(ni$_i$))) ∪ pi; | place |
| 5 | final  node nf level n | ∀ nf$_i$ ∈ NF in level n, Create a place p$_i$ ∈ pg; Pl($\Omega_H$(IODcomp$^{-1}$(nf$_i$))) = Pl($\Omega_H$(IODcomp$^{-1}$(nf$_i$))) ∪ pi; | place |
| 6 | join/fork node jfn | ∀ jfn$_i$ ∈ B in level n, Create an ordinary transition ts$_i$ ∈ pg; Tr_ord($\Omega_H$(IODcomp$^{-1}$(jfn$_i$))) = Tr_ord($\Omega_H$(IODcomp$^{-1}$(jfn$_i$))) ∪ ts$_i$; | transition |
| 7 | merge/decision mdn | ∀ mdn$_i$ ∈ D in level n, Create a place p$_i$ ∈ pg; Pl($\Omega_H$(IODcomp$^{-1}$(mdn$_i$))) = Pl($\Omega_H$(IODcomp$^{-1}$(mdn$_i$))) ∪ p$_i$; | place |
| 8 | connection of the sub-net with the in-Transition | ∀ lf$_i$ ∈ Lf in level n, Create a place p$_{lf}$ ∈ pg; Pl($\Omega_H$(IODcomp$^{-1}$(lf$_i$))) = Pl($\Omega_H$(IODcomp$^{-1}$(lf$_i$))) ∪ p$_{lf}$; Create an arc a = (p$_{lf}$,t$_{in}$) ∈ A; | place + arc |
| 9 | connection of the sub-net with the out-Transition | ∀ lf$_i$ ∈ Lf in level n, Create a place p$_{lf}$ ∈ pg; Pl($\Omega_H$(IODcomp$^{-1}$(mdn$_i$)))=Pl($\Omega_H$(IODcomp$^{-1}$(mdn$_i$))) ∪ p$_{lf}$; Create an arc a$_i$ = (t$_{out}$,p$_{lf}$) ∈ A; | place + arc |
| 10 | IOD edges | See Table II | |

Table II.   Mapping of IOD edges

| Rule | Source node of the transition | Destination node of the transition | $\Omega_H$ (Translation rules) | HCPN |
|------|-------------------------------|-------------------------------------|-------------------------------|------|
| 10.1 | initial or Merge or Decision | Interaction or Join Fork | ∀ e= Ed(i,j) such that (i∈ Ni ∪ D) ∧ (j ∈ I ∪ B); Create an arc a = (Ω(i), Ω(j)) ∈ A; | Arc |
| 10.2 | Initial or Merge Decision | Merge or Decision Final | ∀ e=Ed(i,j) such that (i∈ Ni ∪ D) ∧ (j∈ D ∪ NF) then Create an ordinary transition ts ∈ T; Tr($\Omega_H$(Ref(in)) = Pl($\Omega_H$(Ref(in)) ∪ ts; Create an arc a = (Ω(i),ts) ∈ A; Create an arc a$_i$ = (ts,Ω(j)) ∈ A; | transition + 2 arcs |
| 10.3 | Interaction or Join Fork | Merge or Decision Final | ∀ e=Ed(i,j) such that (i∈ I ∪ B) ∧ (j ∈ D ∪ NF); Create an arc a = (Ω(i),Ω(j)) ∈ A; | Arc |
| 10.4 | Interaction or Join Fork | Interaction or Join Fork | ∀ e=Ed(i,j) such that (i∈ I ∪ B) ∧ (j ∈ I ∪ B) then create a place p ∈ P Pl($\Omega_H$(Ref(in)) = Pl($\Omega_H$(Ref(in))) ∪ p; Create an arc a$_i$ = (Ω(i), p) ∈ A; Create an arc a$_i$ = (p, Ω(j)) ∈ A; | place+ 2 arcs |

Table III shows the transformation of the TD constructs into timed CP-net. For each TD, we make the equivalent timed CP-net construct expressed by an intuitive transformation on as well as a more formal transformation. We call functions and operations used in [3] to make the translation rules like GetMsgOut, GetMsgIn, CreateSeq, GetPointsLF, and MsgSynch. Also, we propose others functions for developing translations rules of TD constructs like Assign and GetColor. Readers can refer this latter reference for more details.

Table III.  Mapping of TD elements

| Rule | TD elements | timed CP-net | |
|------|-------------|--------------|---|
| | | $\theta$ (translation rules) | elements |
| 11 | TD Fragment | Given td ∈ I; Create Page(pg$_i$)∈ Page$_n$; $\theta_1$: I→Page$_n$; $\theta_1$(td)=pg$_i$; | page |
| 12 | Lf | Given lf$_i$ ∈ Lf; Create Place(p$_i$)∈ P; $\theta_2$:Lf→P; $\theta_2$(lf)=p$_i$; | place |
| 13 | Pt | Given pt$_i$ ∈ Pt and If ∈ Lf; i=1;for pt$_i$ = getPointsLf(lf) do createSeq(T$_{lfi-1}$,T$_{lfi}$,P$_{lfi}$) end; if (msgSynch(getMsg(lf, pt$_i$))=true then i::=i+1; end; $\theta_3$:Pt→(P,A,T); $\theta_3$(pt$_i$)=createSeq(T$_{lfi}$,P$_{lfi}$); | place +arc + transition |

| 14 | Msg | Given m ∈ Msg; If₁,If₂ ∈ Lf and p₁,p₂∈EP;m=getMsgOut(If₁,p₁)= getMsgIn(If₂,p₂); createSeq(θ₃(p₁),pₘ,θ₃(p₂)) that θ₃(p₁); θ₃(p₂) ∈ T; θ₄:Msg→P; θ₄(m)=pₘ; | place |
| 15 | State | Given st ∈ State; p ∈ P; Aᵢₙ ∈ A; getColor(p,c) that c(p) ⊂ E; assign(E,Aᵢₙ,p);θ₅:State→P^c;θ₅(st)=p^color; | color in place + arc |
| 16 | Dc | Given dc ∈ Dc and t ∈ T; t= θ₂(getPointsLF(dc)); x=initial time for state ; y=final time for state; Timestamp(t, uniform(x,y)); θ₆:Dc→T^TEMP;θ₆(dc)=t^temp+constraint; | temp in transition output |
| 17 | Tc | Given tc ∈ Tc and t ∈ T; t= θ₂ (getPointsLF(tc)); Timestamp(t, uniform(x,y)); θ₇:Tc→T^G;θ₇(tc)=t^guard+constraint; | guard time in transition |
| 18 | Event | Given e ∈ Event; t∈T; Aᵢₙ∈A; Create place(p ∈P;t= θ₂(getPointsLf(e)); assign(If,p); assign(E,Aₒᵤₜ,p);θ₈:Event→(P,T);θ₈(e)=createSeq(p,t); | place + arc |

## VI. ILLUSTRATED EXAMPLE

In this section, we validate our approach through an example. The considered system is the Automatic Teller Machine (ATM).

### A. Case study modeling

We describe the UML model of the system using IOD. Figure2 shows the class diagram of the ATM system. This diagram describes the static relations between the classes constituting the system, which are System, Client and Bank.

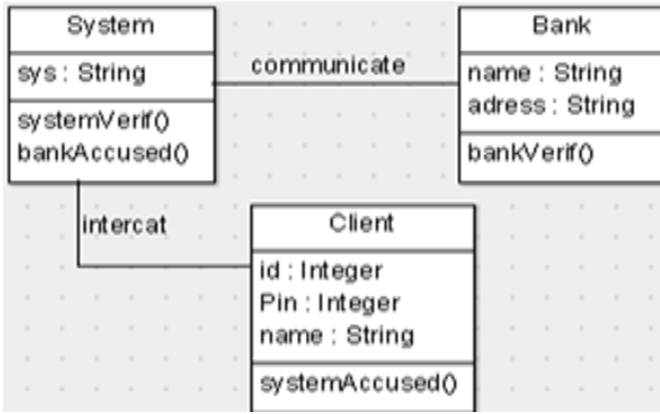

Fig2. UML2 Class Diagram

In order to describe the system behavior and the interactions between the objects, we use the IOD (fig3 shows these interactions). Indeed, the client starts by putting his card into the ATM. After introducing his Personal Identification Number code, the system performs the authentication. When it's valid, the menu is shown. However, in the case of authentication failure, the system performs a new identification. Since the IOD in figure 3 is the main diagram, its hierarchical level is 0.

In a hierarchical way, we refine the behavior of the interaction node Identification (Figure4) through another IOD. In this diagram, three interaction nodes, which are Pin Test, Eject Card and Welcome Message, are identified. Their behavior also is refined, but through SD, as illustrated in figure 5. For clarity reasons, the latter figure shows only the SD associated to the node Pin Test.

### B. Verification of HCPN based model using timed CP-net

After applying the rules mentioned above, a HCPN model is obtained. The figure 6 shows page of obtained model. Using CPN Tools, we can check safety and liveness properties using the Standard ML language. Our work is based on the state space generation. We recall that the final references diagrams should be an SD or a TD. Otherwise, it's not possible to generate a state space. For each given a property, a positive or negative answer is obtained. However, our system is resettable. Additional properties may be checked based on the work proposed in [12].

## VII. CONCLUSION AND FUTURE WORK

This work aimed at the formalization of the hierarchical use of IOD, in order to formally validate the models based on these diagrams. For this purpose, first, we recalled previous work. Then, we proposed the IOD formalization in hierarchical way. Finally, we created the mapping function that maps a model constituted of IOD, SD and TD into a HCPN model using timed CP-net. Using CPNTools, the formal verification of supported properties can be performed. The obtained results are promising. A tool for the automatic generation of our approach is currently under development. As future work, we propose to apply our approach to an industrial system by using specifics properties.

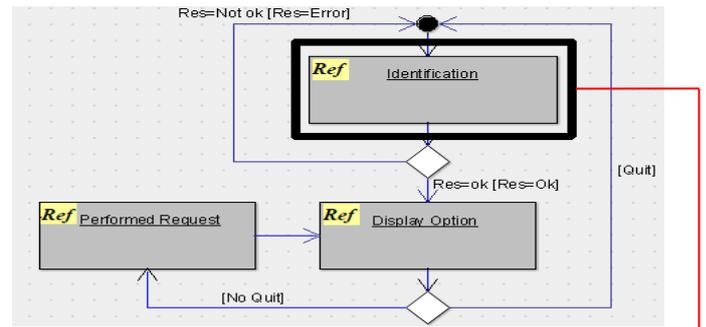

Fig3. IOD main (level0)

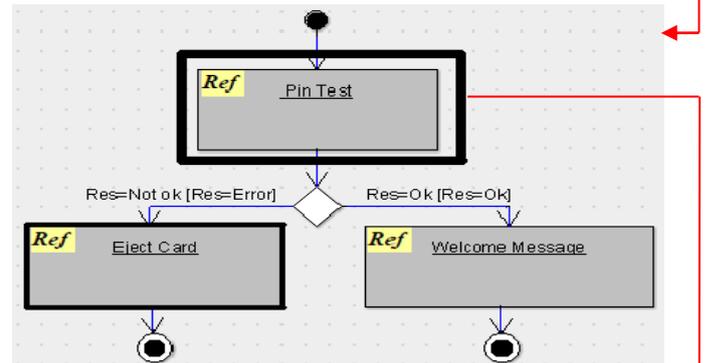

Fig4. IOD level1 (Identification Interaction node)

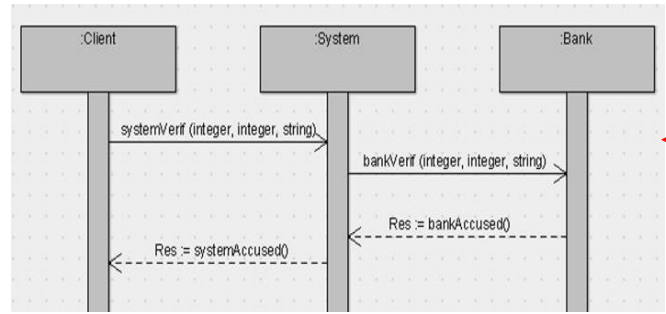

Fig5. Pin Test SD

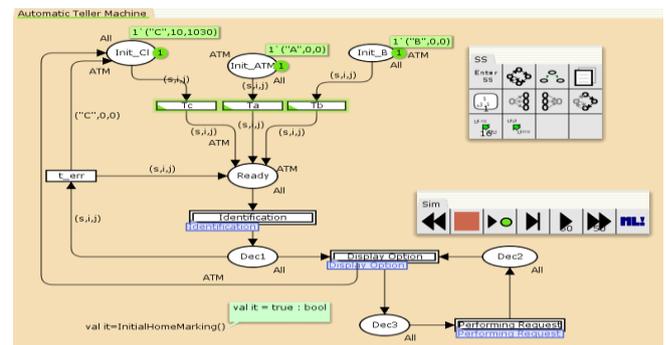

Fig6. HCPN prime page derived from IOD level 0